\documentclass[journal]{IEEEtran}
\usepackage{amsmath,graphicx}
\usepackage{enumerate}
\usepackage{graphics}
\usepackage{amsbsy}
\usepackage{amsmath}
\usepackage{amssymb}
\usepackage{amsthm}
\usepackage{amscd}
\usepackage{subfigure}
\usepackage{color}
\usepackage{algorithm}
\usepackage{algorithmic}
\usepackage{epstopdf}
\usepackage{pgf,tikz}
\usetikzlibrary{arrows}

\newcommand{\argmax}{\operatornamewithlimits{argmax}}

\newtheorem*{Theorem1}{Theorem 1}

\newtheorem*{Lemma1}{Lemma 1}
\newtheorem*{Lemma2}{Lemma 2}

\title{A new structure exploiting derivation\\ of recursive direct weight optimization }
\author{Liang Dai and Thomas B. Sch\"{o}n
\thanks{This work is supported by the Swedish research Council (VR) project with number 106400801. The authors are with the Department of Information Technology, Uppsala University, SE 751 05, Uppsala, Sweden.}
}
\date{}
\begin{document}
\definecolor{xdxdff}{rgb}{0.49,0.49,1}
\definecolor{qqqqff}{rgb}{0,0,1}
\maketitle
\begin{abstract}
The recursive direct weight optimization method is used to solve challenging nonlinear system identification problems. This note provides a new derivation and a new interpretation of the method. The key underlying the note is to acknowledge and exploit a certain structure inherent in the problem.
\end{abstract}

\section{Introduction}
\IEEEPARstart{N}{onlinear} system identification is a very active and diverse research field. In the past, many approaches have been suggested and tested. For a nice survey of some of these results, please refer to the special issue \cite{Ljung} and the recent edited book \cite{bai_bk}. Direct Weight Optimization (DWO) is one of these approaches which is obtained by forming a direct linear combination of the outputs, and finding the corresponding coefficients by optimizing a suitably chosen criteria. For more details, please refer to \cite{dwo1,dwo2,dwo3}.

In \cite{bai}, the authors proposed the so called Recursive Direct Weight Optimization (RDWO) approach to solve the same problem in a recursive fashion. The idea of such approach is based on minimizing the probability that an upper bound of the estimation error is larger than a predefined threshold.

Note that we are not intending to give an improvement over the RDWO method, instead we are aiming at providing a derivation and interpretation of the RDWO method that is natural and direct. More specifically, the contributions of this note lie in the following two points: 1) A new and simpler derivation of the method is given; 2) A novel interpretation of the RDWO method is provided. The key allowing for this is the newly introduced quantities  $\{\hat{\varphi}_{x}(k)\}_{k=1}^{N}$ defined in the subsequent section, which exploit the inherent structure information of the method.

\section{The RDWO method}

Let us start by a brief review of the RDWO method, following the notations introduced in \cite{bai}. The nonlinear system considered is given by
\begin{align}
y(k) = f(\varphi(k)) + e(k),\qquad  k = 1,\cdots,N,
\end{align}
where the nonlinear function $f(\cdot)$ is assumed to be differentiable with Lipschitz constant $L_1$, i.e.
 \begin{align*}
 \left|\frac{df(\varphi)}{d\varphi}\right|\le L_1,\ \varphi \in \mathbb{R}.
 \end{align*}
 For $k = 1,\cdots,N$, $y(k)\in \mathbb{R}$ are observations, $\varphi(k)\in \mathbb{R}$ are inputs, and $e(k)\in \mathbb{R}$ are noise terms, which are assumed to be independently identically distributed (i.i.d.) Gaussian noise with zero mean and variance $\sigma_e^2$.

 The estimate $\hat{f}(x)$ is defined as
\begin{align}
\hat{f}(x) = \sum_{k=1}^{N} w_x(k)y(k), \label{eq.dwo}
\end{align}
where $\sum_{k=1}^{N}w_x(k) = 1.$

For a given $x$, define $\widetilde{\varphi}_x(k) = |x - \varphi(k)|$, for $k = 1,\cdots,N$. The following upper bound of the squared estimation error $(\hat{f}(x) - f(x))^2$ can be obtained:
\begin{align}
(\hat{f}(x) - f(x))^2 \le z^2,
\end{align}
where
\begin{align*}
z =L_1\sum_{k=1}^N |w_x(k)|\widetilde{\varphi}_x(k) + |\sum_{k=1}^{N}w_x(k)e(k)|.
\end{align*}
In \cite{bai} it is mentioned that this bound is in fact tight in the sense that there exist functions and noise distributions for which the inequality becomes equality.

The RDWO method is to minimize the probability when $z$ is greater than a predefined $\delta'$, that is to solve
\begin{align}
\min_{w_x(k)}\text{Prob}(z\ge \delta'), \quad \text{s.t.} \ \sum_{k=1}^N w_x(k) =1,
\end{align}
where $\delta'$ satisfies $\delta' > L_1\sum_{k=1}^N |w_x(k)|\widetilde{\varphi}_x(k).$

Since the noise terms are assumed to be i.i.d Gaussian distributed, $\text{Prob}(z\ge \delta')$ can be explicitly calculated. For the details, please refer to \cite{bai}. As such, it turns out that the optimized weights can be obtained by solving the following optimization problem.\\

For given $\delta>0$, $N$, $x \in \mathbb{R}$, solve:
\begin{align}
\widehat{w}_x=&\argmax_{w_{x}(k)} \frac{\delta - \sum_{k=1}^{N}|w_x(k)|\widetilde{\varphi}_x(k)}{\sqrt{\sum_{k=1}^{N}w_x(k)^2}}\label{problem} \\
\text{s.t.}\quad & \sum_{k=1}^{N} w_x(k) =1,\nonumber
\end{align}
in which $\delta = \frac{\delta'}{L_1}.$ The main result in \cite{bai} is phrased as follows, which gives the analytical solution to the problem ($\ref{problem}$).
\begin{Theorem1}\label{Theorem 1}
Suppose that $\delta > \min_{1 \leq k \leq N}\widetilde{\varphi}_x(k)$. Let $M_x \triangleq \{ m_1,m_2, \cdots,m_l \}$ be a set such that $m \in M_x \Leftrightarrow \delta > \widetilde{\varphi}_x(m)$. Then the solution to ($\ref{problem}$) is unique and given by
\begin{equation} \label{orgsol}
\widehat{w}_x(k) = \left\{
\begin{matrix}
\frac{\delta - \widetilde{\varphi}_x(k)}{l\delta-\sum_{i=1}^{l}\widetilde{\varphi}_x(m_i)},&k\in M_x,\\
0,   &k\notin M_x.\\
\end{matrix}\right.
\end{equation}
\end{Theorem1}

Based on Theorem 1, a recursive scheme, i.e. the RDWO method, for updating the weights in (\ref{eq.dwo}) is derived when new observations are obtained.

\section{Main results}
\subsection{Derivation}
Lemma 1 in \cite{bai} will be reused for our analysis, which is cited as follows.
\begin{Lemma1}
The problem given in ($\ref{problem}$) is equivalent to the following optimization problem: For given $\delta>0$, $N$, $x \in \mathbb{R}$, solve:
\begin{align}
\widehat{w}_x=&\argmax_{w_{x}} \frac{\delta - \sum_{k=1}^{N}w_x(k)\widetilde{\varphi}_x(k)}{\sqrt{\sum_{k=1}^{N}w_x(k)^2}}\label{problem1} \\
\mathrm{s.t.} \quad & \sum_{k=1}^{N} w_x(k) =1 \nonumber\\
& w_x(k) \ge 0  \nonumber
\end{align}
\end{Lemma1}

Let us start by introducing the following new key quantities,
\begin{align}
\widehat{\varphi}_x(k)= \delta-\widetilde{\varphi}_x(k), \qquad \forall \ k = 1,\cdots, N.
\label{eq.def1}
\end{align}

We remark that different from \cite{bai}, the results given in this note rely heavily on the quantities $\{\widehat{\varphi}_{x}(k)\}_{k=1}^{N}$ introduced in (\ref{eq.def1}). By introducing them, the derivation and the interpretation of the RDWO method become simpler. The reason is that $\{\widehat{\varphi}_{x}(k)\}_{k=1}^{N}$ reveals a particular structure inherent in the problem.

A useful relation relating the different definitions introduced so far is given by
\begin{align}
m \in M_x\, \Leftrightarrow\, \delta > \widetilde{\varphi}_x(m)\, \Leftrightarrow\, \widehat{\varphi}_x(m)>0.
\label{def}
 \end{align}

 Notice the following fact
\begin{align*}
\delta - |x-\varphi(k)| = \min\{\varphi(k)- (x-\delta), (x+\delta)-\varphi(k)\},
\end{align*}
so when $k \in M_{x}$ , i.e. when $\phi(k)$ lies in the interval $(x-\delta,x+\delta)$, the values $\varphi(k)- (x-\delta)$ and $(x+\delta)-\varphi(k)$ measure the distances between $\varphi(k)$ and the points $x-\delta$ and $x+\delta$ separately. Hence,  $\widehat{\varphi}_x(k)$ measures the distance between $\varphi(k)$ and the set of the endpoints of $(x-\delta,x+\delta)$. An illustration of $\widehat{\varphi}_x(k)$ is given in Fig. \ref{fig.uqp}.

\begin{figure}[htbp] 
   \centering
\begin{tikzpicture}[line cap=round,line join=round,>=triangle 45,x=1.0cm,y=1.0cm]
\clip(6,5) rectangle (15,11.5);
\draw [line width=0.4pt] (7,10)-- (13.66,10.02);
\draw (11.84,10.01)-- (13.66,10.02);
\draw [line width=1.5pt] [color=qqqqff](11.84,10.01)-- (13.66,10.02);
\draw (12.44,10.76) node[anchor=north west] {$\widehat{\varphi}_x(k)$};
\draw [line width=0.4pt] (7,7)-- (13.66,7.02);
\draw (8.82,7.01)-- (13.66,7.02);
\draw (7.54,7.84) node[anchor=north west] {$\widehat{\varphi}_x(k)$};
\draw [line width=1.5pt] [color=qqqqff] (7,7)-- (8.82,7.01);
\draw (9.9,8.96) node[anchor=north west] {(a)};
\draw (9.98,5.64) node[anchor=north west] {(b)};
\begin{scriptsize}
\fill  (7,10) circle (1.7pt);
\draw (7.22,9.7) node {$x-\delta$};
\fill   (13.66,10.02) circle (1.7pt);
\draw  (13.88,9.62) node {$x+\delta$};
\fill  (10.34,10.01) circle (1.7pt);
\draw (10.32,9.66) node {$x$};
\fill  (11.84,10.01) circle (1.7pt);
\draw (11.96,9.62) node {$\varphi(k)$};
\fill (7,7) circle (1.5pt);
\draw (7.22,6.7) node {$x-\delta$};
\fill (13.66,7.02) circle (1.7pt);
\draw (13.88,6.62) node {$x+\delta$};
\fill (10.34,7.01) circle (1.7pt);
\draw (10.32,6.66) node {$x$};
\fill (8.82,7.01) circle (1.7pt);
\draw (8.94,6.6) node {$\varphi(k)$};
\end{scriptsize}
\end{tikzpicture}
   \caption{This figure illustrates the meaning of $\widehat{\varphi}_x(k)$ when $k \in M_{x}$ in different cases. We can see that, $\widehat{\varphi}_x(k)$ measures the distance between $\varphi(k)$ and the set of endpoints of interval $(x-\delta,x+\delta)$ which is illustrated by the lengths of the thick blue segments. }
   \label{fig.uqp}
\end{figure}

Next, we will transform the problem in (\ref{problem1}) into an equivalent formulation using $\widehat{\varphi}_x(k)$. Notice that since
 \begin{align*}\sum_{k=1}^{N} w_x(k) =1,\end{align*}
we have
  \begin{align*}
  \delta - \sum_{k=1}^{N}w_x(k)\widetilde{\varphi}_x(k) &= \delta\sum_{k=1}^{N}w_x(k) - \sum_{k=1}^{N}w_x(k)\widetilde{\varphi}_x(k) \\
  &= \sum_{k=1}^{N}w_x(k)\widehat{\varphi}_x(k),
  \end{align*}
  which implies that the problem (\ref{problem1}) can be rewritten as
\begin{align}\label{problem2}
\widehat{w}_x=&\argmax_{w_{x}} \frac{ \sum_{k=1}^{N}w_x(k)\widehat{\varphi}_x(k)}{\sqrt{\sum_{k=1}^{N}w_x(k)^2}} \\ \nonumber
\mathrm{s.t.}\quad &\sum_{k=1}^{N} w_x(k) =1 \\ \nonumber
  &w_x(k) \ge 0 \nonumber
\end{align}

The sketch of the following part is as follows. Based on the new formulation in (\ref{problem2}), we will establish a lemma, which makes it possible to determine the zero elements of $\widehat{w}_x$ before actually solving the problem (\ref{problem2}). When the zero elements of $\widehat{w}_x$ have been identified beforehand, the remaining task is to determine those nonzero elements. These nonzero elements are obtained by application of the Cauchy-Schwartz inequality.

\begin{Lemma2}
For the problem given in ($\ref{problem2}$), if $\widehat{\varphi}_x(k) \leq 0$, then it holds that $\widehat{w}_x(k)=0$.
\end{Lemma2}

\begin{IEEEproof}
The proof is given by a contradiction argument. Assume that $\widehat{\varphi}_x(k) \leq 0$, but $\widehat{w}_x(k) \neq 0$.

First, we show that $\widehat{w}_x(k) < 1$ holds. According to the assumptions given in Theorem 1, we have that $$\delta > \min_{1 \leq j \leq N}\widetilde{\varphi}_x(j),$$  which gives that the maximum value of the objective function in (\ref{problem1}) and also in (\ref{problem2}) will be positive. Together with the fact that $\sum_{1 \leq j \leq N}\widehat{w}_x(j) = 1$, one has that $\widehat{w}_x(k) \neq 1$, otherwise the maximum value of the objective function in (\ref{problem2}) will be non-positive.

Since $\widehat{w}_x(k) \neq 1$, we can construct another point $\bar{w}_x$ satisfying the constraints to ($\ref{problem2}$), which is given as:
\begin{align}
\bar{w}_x(i) = \left\{
\begin{matrix}
&\frac{\widehat{w}_x(i)}{1-\widehat{w}_x(k)}, &i \neq k,\\
&0,   &i = k,
\end{matrix}\right.\nonumber
\end{align}
for $i = 1,\cdots,N.$

It will be proven shortly that the objective function will increase at the point $\bar{w}_x$, which contradicts the fact that $\widehat{w}_x$ is the optimal value, thus the proof is concluded.

The reasoning is given by the following arguments.
\begin{align}
\frac{\sum_{i=1}^{N}\bar{w}_x(i)\widehat{\varphi}_x(i)}{\sqrt{\sum_{i=1}^{N}\bar{w}_x(i)^2}} \,= \,&\frac{\sum_{i=1,i\neq k}^{N}\frac{\widehat{w}_x(i)}{1-\widehat{w}_x(k)}\widehat{\varphi}_x(i)}{\sqrt{\sum_{i=1,i \neq k}^{N}(\frac{\widehat{w}_x(i)}{1-\widehat{w}_x(k)})^2}} \nonumber \\
=\, &\frac{\sum_{i=1,i \neq k}^{N}\widehat{w}_x(i)\widehat{\varphi}_x(i)}{\sqrt{\sum_{i=1,i \neq k}^{N}\widehat{w}_x(i)^2}} \nonumber \\
\stackrel{(I)} >\, &\frac{\sum_{i=1}^{N}\widehat{w}_x(i)\widehat{\varphi}_x(i)}{\sqrt{\sum_{i=1}^{N}\widehat{w}_x(i)^2}}. \label{eq.seq}
\end{align}
The inequality $(I)$ follows from the following facts. By assumption,  $\widehat{\varphi}_x(k)\le 0$ and $\widehat{w}_x(k) > 0$ hold, so we have
\begin{align*}
\sum_{i=1,i \neq k}^{N}\widehat{w}_x(i)\widehat{\varphi}_x(i) \ge \sum_{i=1}^{N}\widehat{w}_x(i)\widehat{\varphi}_x(i).
\end{align*}

Also, since $\widehat{w}_x(k) \ne 0$, it follows that
\begin{align*}
0<\sqrt{\sum_{i=1,i \neq k}^{N}\widehat{w}_x(i)^2} < \sqrt{\sum_{i=1}^{N}\widehat{w}_x(i)^2}.
\end{align*}

These two facts conclude inequality $(I)$, and finalize the proof.
\end{IEEEproof}

From Lemma 2, we can conclude that to optimize (\ref{problem2}), one only needs to optimize over the weights for those $\widehat{\varphi}_x(i)$ which are positive. With the facts given in Lemma 2, the remaining steps for the derivation of Theorem 1 are as follows.
\begin{IEEEproof}
The optimization problem (\ref{problem2}) can be translated into the following problem:
\begin{align*}
\widehat{w}_x=&\argmax_{w_{x}(k)} \frac{ \sum_{k\in M_x}w_x(k)\widehat{\varphi}_x(k)}{\sqrt{\sum_{k\in M_x}w_x(k)^2}} \\
\mathrm{s.t.}\quad &\sum_{k\in M_x} w_x(k) =1 \\
  &w_x(k) \geq 0
\end{align*}

Applying the Cauchy-Schwartz inequality, we have that
\begin{align}
&\,\frac{ \sum_{k\in M_x}w_x(k)\widehat{\varphi}_x(k)}{\sqrt{\sum_{k\in M_x}w_x(k)^2}} \nonumber \\
\leq\,  &\frac{\sqrt{\sum_{k\in M_x}w_x^2(k)}\sqrt{\sum_{k\in M_x}\widehat{\varphi}_x^2(k)}}{\sqrt{\sum_{k\in M_x}w_x(k)^2}} \nonumber \\
=\, & \sqrt{\sum_{k\in M_x}\widehat{\varphi}_x(k)^2}, \label{cauchy}
\end{align}
and the '=' holds when $\frac{w_x(k)}{\widehat{\varphi}_x(k)}= C,k\in M_x$, in which $C$ is a constant which will be determined shortly.

Since $\sum_{k\in M_x}w_x(k)=1$, we have that $\sum_{k\in M_x}C \widehat{\varphi}_x(k)=1$, which in turn gives that $$C = \frac{1}{\sum_{k\in M_x}\widehat{\varphi}_x(k)}.$$

Finally we conclude that
\begin{equation}\label{newsol}
\widehat{w}_x(k) = \left\{
\begin{matrix}

\frac{\widehat{\varphi}_x(k)}{\sum_{i \in M_x}\widehat{\varphi}_x(m_i)},&k\in M_x,\\
0,   &k\notin M_x.\\
\end{matrix}\right.
\end{equation}
\end{IEEEproof}
We end this section by remarking that the formulation in (\ref{newsol}) is equivalent to the formulation in (\ref{orgsol}), which easily follows from the definitions in ($\ref{eq.def1}$).

\subsection{Interpretation}
In this section, we will give a novel interpretation of the RDWO method based on $\widehat{\varphi}_x(k)$ introduced in ($\ref{eq.def1}$).

In order to describe the recursive algorithm, the following time-dependent sets are used. For a given $x$, $M_{x}^{N}$ is defined as the index set for the inputs $\{ \varphi(k)\}_{k=1}^{N}$ which lie in the interval $(x-\delta,x+\delta)$,  and $\{w_{k}^{N}(x)\}_{k=1}^{N}$ are the weights obtained by the RDWO method when $N$ observations are obtained, $f_{N}(x)$ is the approximated function value at time $N$.

Based on the formulation in (\ref{newsol}), the RDWO algorithm can be interpreted as follows. At time $N+1$, for a given $x$, when a new observation is obtained, calculate the distance from its corresponding input to $x$. If the distance is greater than $\delta$, then keep the previous weights unchanged and assign no weight for the current observation; otherwise reweigh all the weights at time $N$ by a factor $\lambda_{N+1}$, which will be defined in (\ref{ratio}), and assign the current observation with weight $1 - \lambda_{N+1}.$  The factor $\lambda_{N+1}$ measures the ratio between the total distances of the inputs in $M_{x}^{N}$ to the endpoints of interval $(x-\delta,x+\delta)$ and the total distances of the inputs in $M_{x}^{N+1}$ to the endpoints of interval $(x-\delta,x+\delta)$.

In conclusion, the new formulation of the RDWO algorithm is summarized in Algorithm \ref{alg.rdwo}.
\begin{algorithm}[]
  \caption{Recursive Direct Weight Optimization}
  \label{alg.rdwo}
  \begin{algorithmic}[1]
    \STATE Collect new data $y(N+1)$, $\varphi(N+1)$.
    \STATE Calculate $$\widehat{\varphi}_x(N+1) = \delta - |x- \varphi(N+1)|.$$
    \IF{$\widehat{\varphi}_x(N+1) \le 0$} \nonumber
    \vspace{+1mm}
        \STATE Set $M^{N+1}_x = M^{N}_x.$ \nonumber
        \vspace{+1mm}
        \STATE Update $w_k^{N+1}$ according to
                        \begin{align*}
                         w_k^{N+1}(x) =
                          \begin{cases}
                           w_k^{N}(x), & \text{if } k =1,2,\cdots, N, \\
                           0,       & k = N+1.
                          \end{cases}
                         \end{align*}
        \STATE Set $f_{N+1}(x) = f_N(x).$ \nonumber
        \STATE Set $N \leftarrow N+1$, and go back to iterate from step 1.
    \ELSE
        \STATE Set $M^{N+1}_x = M^{N}_x + m_{l+1},$ where $m_{l+1} = N+1.$
        \STATE Calculate
               \begin{align}
               \label{ratio}
               \lambda_{N+1} =\frac{\sum_{j=1}^{l}\widehat{\varphi}_x(m_j)}{\sum_{j=1}^{l+1} \widehat{\varphi}_x(m_j)},
               \end{align}
        \STATE Update $w_k^{N+1}$ according to
                 \begin{align*}
                 w_k^{N+1}(x) =
                  \begin{cases}
                   \lambda_{N+1} w_k^{N}(x),  & \text{if } k =1,2,\cdots, N, \\
                   1-\lambda_{N+1} ,     & k = N+1.
                  \end{cases}
                 \end{align*}
        \STATE Update $f_{N+1}(x)$ according to
                \begin{align*}
                f_{N+1}(x) = \lambda_{N+1}f_N(x) + (1-\lambda_{N+1})y(N+1).
                \end{align*}
        \STATE Set $l \leftarrow l+1$ and $N \leftarrow N+1$, and go back to iterate from step 1.
    \ENDIF
  \end{algorithmic}
\end{algorithm}

\section{Conclusion}
This note first presents a novel derivation of the recursive direct weight optimization algorithm by introducing new quantities which can exploit useful structure information inherent in the problem. Based on the formulation provided by the new derivation, a new interpretation of the algorithm is also obtained. We end the discussions by remarking the following two points: 1) The studies about the consistency properties and other related issues of the RDWO in \cite{bai} are also valid to the new formulation derived in the note, since the two formulations are mathematically equivalent. 2) The difference compared to the earlier result in \cite{bai} is that by introducing the new structure exploiting quantities, the derivation and the interpretation of the RDWO are made more transparent.

\section{Acknowledgement}
The authors would like to thank Johannes Nygren for useful comments on an early draft of the note.

\end{document}